\documentclass[12pt]{article}
\usepackage{amsmath}
\usepackage{graphicx}

\usepackage{geometry}
\usepackage[tight]{subfigure}
\ifx\pdfoutput\@undefined\usepackage[usenames,dvips]{color}
\else\usepackage[usenames,dvipsnames]{color}
\IfFileExists{pdfcolmk.sty}{\usepackage{pdfcolmk}}{} 
\fi
\usepackage[plainpages=false,pdfpagelabels,pagebackref=false,naturalnames=true,hyperindex=true,pdftitle={The Sigma Profile},pdfauthor={Carlos Gershenson}]{hyperref}
\hypersetup{colorlinks=true,
urlcolor=Cerulean,linkcolor=BrickRed,citecolor=RoyalBlue,a4paper,
  pdfpagemode=None,
  pdfstartview=FitH}

\begin{document}

\title{The Sigma Profile: A Formal Tool to Study Organization and its Evolution at Multiple Scales}
\author{Carlos Gershenson$^{1,2}$ \\
$^{1}$ New England Complex Systems Institute\\
24 Mt. Auburn St. Cambridge MA 02138, USA\\
\href{mailto:carlos@necsi.org}{carlos@necsi.org} \ \url{http://necsi.org} \\
$^{2}$Centrum Leo Apostel, Vrije Universiteit Brussel\\
Krijgskundestraat 33 B-1160 Brussel, Belgium\\
\href{mailto:cgershen@vub.ac.be}{cgershen@vub.ac.be} \ \url{http://homepages.vub.ac.be/~cgershen}}
\maketitle

\begin{abstract}
The $\sigma$ profile is presented as a tool to analyze the organization of systems at different scales, and how this organization changes in time.  Describing structures at different scales as goal-oriented agents, one can define $\sigma \in [0,1]$ ("satisfaction") as the degree to which the goals of each agent at each scale have been met. $\sigma$ reflects the organization degree at that scale. The $\sigma$ profile of a system shows the satisfaction at different scales, with the possibility to study their dependencies and evolution.
It can also be used to extend game theoretic models. 
A general tendency on the evolution of complexity and cooperation naturally follows from the $\sigma$ profile.
Experiments on a virtual ecosystem are used as illustration.

\end{abstract}

\section{Introduction}

We use metaphors, models, and languages to describe our world. Different descriptions may be more suitable than others. We tend to select from a pool of different descriptions those that fit with a particular purpose. Thus, it is natural that there will be several useful, overlapping descriptions of the same phenomena, useful for different purposes.

In this paper, the $\sigma$ profile is introduced to describe the organization of systems at multiple scales. Some concepts were originally developed for engineering \cite{GershensonDCSOS}. Here they are extended with the purpose of scientific description, in particular to study evolution.

This article is organized as follows. In the following sections, concepts from multi-agent systems, game theory, and multiscale analysis are introduced. These are then used to describe natural systems and evolution. In the following sections, a simple simulation and experiments are presented to illustrate the $\sigma$ profile. Conclusions close the paper.

\section{Agents}

Any phenomenon can be \emph{described} as an agent. An agent is a description of an entity that \emph{acts} on its environment \cite[p. 39]{GershensonDCSOS}.
Thus, the terminology of  multi-agent systems \cite{Maes1994,WooldridgeJennings1995,Wooldridge2002,Schweitzer2003} can be used to describe complex systems and their elements. An electron acts on its surroundings with its electromagnetic field, a herd acts on an ecosystem, a car acts on city traffic, a company acts on a market. Moreover, an observer can ascribe \emph{goals} to an agent. An electron tries to reach a state of minimum energy, a herd tries to survive, a car tries to get to its destination as fast as possible, a company tries to make money. We can define a variable $%
\sigma$ to represent \emph{satisfaction}, i.e. the degree to which the goals of an agent have been reached. This will also reflect the organization of the agent \cite{Ashby1962,GershensonHeylighen2003a,GershensonDCSOS}.

Since agents act on their environment, they can affect positively, negatively, or neutrally the satisfaction of other agents. We can define the \emph{friction} $\phi _{i}$ between agents $A$ and $B$ as

\begin{equation}
\phi _{A,B}=\frac{-\Delta \sigma _{A}-\Delta \sigma _{B} }{2%
}.  \label{eqFrictionAB}
\end{equation}

This implies that when the decrease in $\Delta \sigma$ (satisfaction reduced) for one agent is greater than the increase in $\Delta \sigma$ (satisfaction increased) for the other agent, $\phi _{A,B}$ will be positive. In other words, the satisfaction gain for one agent is lesser than the loss of satisfaction in the other agent. The opposite situation, i.e. a negative $\phi _{A,B}$ implies an overall increase in satisfaction, i.e. synergy.

Generalizing, the friction within a group of $n$ agents will be

\begin{equation}
\phi_{n}=\frac{-\sum_{i}^n \Delta \sigma _{i}}{n%
}.  \label{eqFrictionn}
\end{equation}

Satisfactions at different scales can also be compared. This can be used to study how satisfaction changes of elements affect satisfaction changes of the system they compose.

\begin{equation}
\phi _{n,sys}=\frac{\phi_{n}-\Delta \sigma _{sys} }{2%
}.  \label{eqFrictionnsys}
\end{equation}

\section{Games}

Game theory \cite{vonNeumannMorgenstern1944,JMS1982} and in particular the prisoner's dilemma \cite{Tucker:1950,Poundstone:1992} have been used to study mathematically the evolution of cooperation \cite{Axelrod1984}. It will be used here to exemplify the concepts of the $\sigma$ profile. A well studied abstraction is given when players (agents) choose between cooperation or defection. A cooperator pays a cost $c$ for another to receive a benefit $b$, while a defector does not pay a cost nor deals benefits \cite{Nowak2006} \footnote{$b>c$ is assumed.}. The possible interactions can be arranged in the two by two matrix (\ref{eq:bcmatrix}), where the payoff refers to the `row player' $A$. When both cooperate, $A$ pays a cost ($-c$), but receives a benefit $b$ from $B$. When $B$ defects, $A$ receives no benefit, so it loses $-c$. This might tempt $A$ to defect, since it will gain $b>b-c$ if $B$ cooperates and will not lose if $B$ also defects as $0>-c$.

\begin{equation}
\label{eq:bcmatrix}
 \begin{array}{ccc}
  A\backslash B & C & D \\
   C & b-c & -c  \\
   D & b & 0 
\end{array}
\end{equation}

We can use the payoff of an agent to measure its satisfaction $\sigma$. Moreover, we can calculate the friction between agents $A$ and $B$ with $\phi _{A,B}$ (\ref{eqFrictionAB}) as shown in (\ref{eq:frictionMatrix}):

\begin{equation}
\label{eq:frictionMatrix}
 \begin{array}{ccc}
  \phi _{A,B} & C & D \\
   C & -(b-c) & -\frac{b-c}{2}  \\
   D & -\frac{b-c}{2} & 0 
\end{array}
\end{equation}

If we assume that $A$ and $B$ form of a system, we can define naively its satisfaction as the sum of the satisfactions of the elements. Therefore, the satisfaction of the system $\sigma_{A,B}$ would be also the negative of (\ref{eq:frictionMatrix}) times two:

\begin{equation}
\label{eq:sigmaSysMatrix}
 \begin{array}{ccc}
  \sigma_{A,B} & C & D \\
   C & 2(b-c) & b-c  \\
   D & b-c & 0 
\end{array}
\end{equation}

Thus, we can study satisfactions at two different scales. At the lower scale, agents are better off defecting, given the conditions of (\ref{eq:bcmatrix}). However, at the higher scale, the system will have a higher satisfaction if agents cooperate, since $b-c>\frac{b-c}{2}>0$. Here we can see that \emph{reducing the friction $\phi$ at the lower scale increases the satisfaction $\sigma_{sys}$ at the higher scale}. This has been shown to be valid also for the more general case, when $\sigma_{sys}\neq \sum_{i}^n \sigma _{i}$ and has been used as a design principle to engineer self-organizing systems \cite{GershensonDCSOS}.

\section{Multiscale Analysis}

Bar-Yam proposed multiscale analysis \cite{BarYam2004} to study the complexity of systems as scale varies. In particular, one can visualize this with the ``complexity profile" of a system, i.e. the complexity of a system depending on the scale at which it is described. Here complexity is understood as the amount of information required to describe a system \cite{Prokopenko:2008}.

\begin{figure}
\begin{center}
\includegraphics[width=0.6\textwidth]{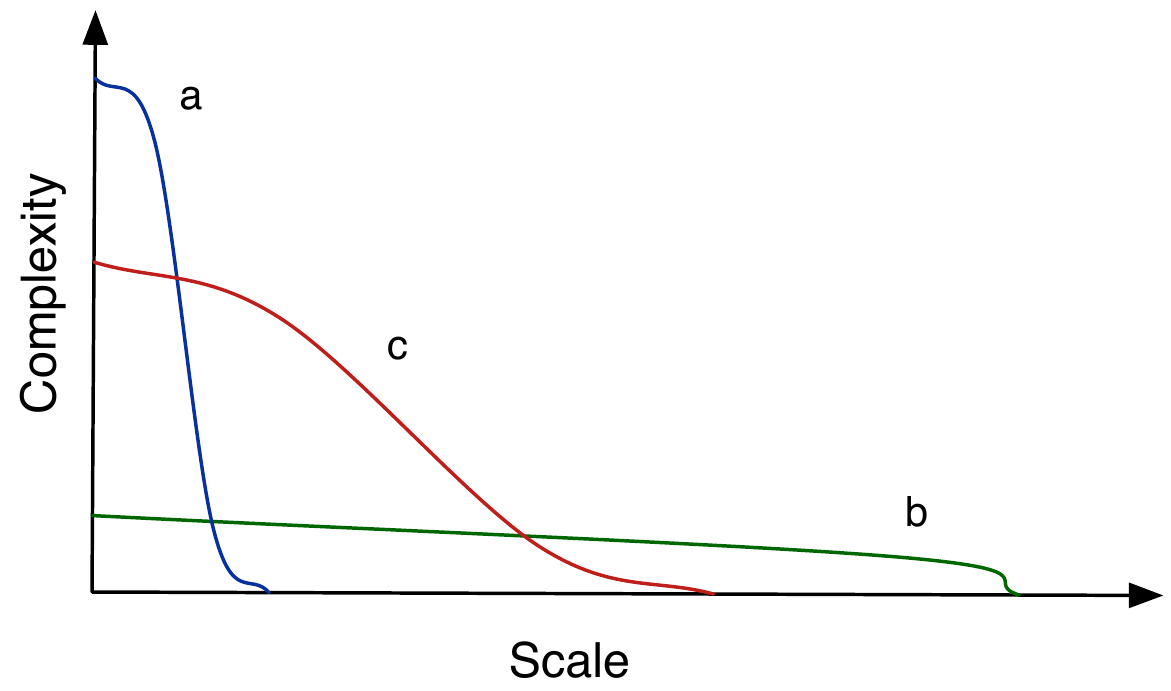}
\caption{Complexity profile for three systems. See text for details.}
\label{fig:cxProfile}
\end{center}
\end{figure}

As an example, Figure \ref{fig:cxProfile} shows the complexity profile of three systems: Curve $a$ represents 1mm$^3$ of a gas. At a low (atomic) scale, its complexity is high, since one needs to specify the properties of each atom to describe the system. However, at higher scales these properties are averaged, so the complexity is low. Curve $b$ represents 1mm$^3$ of a comet. Since the atoms are stable, few information is required to describe the system at low scales, since these are relatively regular. Still, as the comet travels large distances, information is relevant at very high scales. Curve $c$ represents 1mm$^3$ of an animal. Its atoms are more ordered as $a$, but less than $b$, so its complexity at that scale is intermediate. Given the organization of living systems, the complexity required to describe $c$ at the mesoscale is high. For high scales, however, the complexity of $b$ is higher, since the information of $c$ is averaged with that of the rest of the planet.

Generalizing Ashby's law of requisite variety \cite{Ashby1956}, multiscale analysis can be used to match the complexity of a system at different scales to an appropriate control method. This is because systems are doomed to fail when their complexity does not match the complexity of their environment. In other words, solutions need to match the complexity of the problem they are trying to solve at a particular scale.

\begin{figure}[htbp]
\begin{center}
\includegraphics[width=0.6\textwidth]{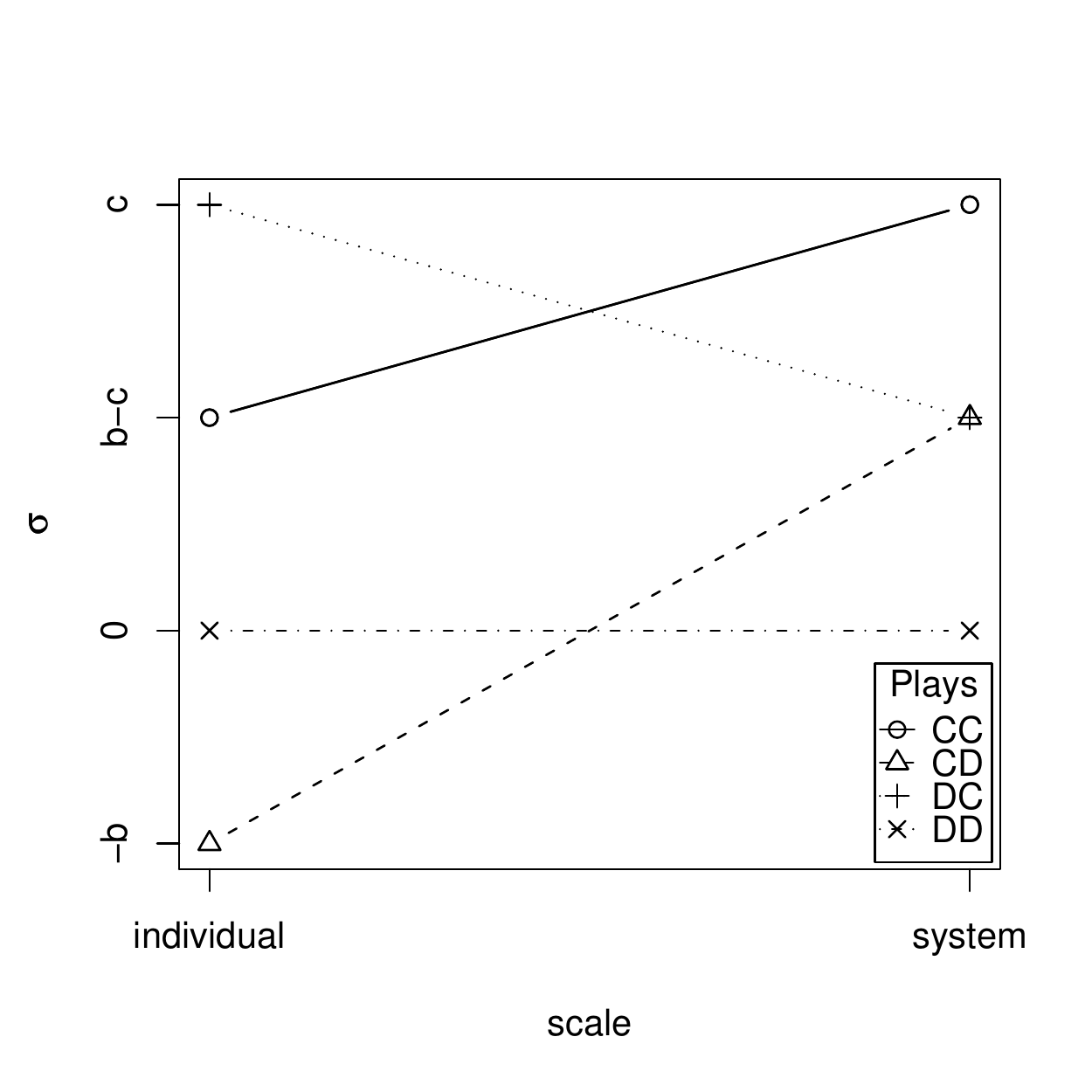}
\caption{$\sigma$ profile for the prisoner's dilemma. At the individual scale, the best play is DC, while at the system level it is CC. For graphical purposes, $b=2c$.}
\label{fig:sigmaPD}
\end{center}
\end{figure}

Inspired by the complexity profile, the $\sigma$ profile is the comparison of satisfaction according to scale. Figure \ref{fig:sigmaPD} shows the $\sigma$ profile for the prisoner's dilemma example described above. There are two scales on the $x$ axis: individual and system. The $y$ axis indicates the satisfaction at different levels, for different combinations of two players choosing between cooperation (C) and defection (D). For the individual scale, the play combination that gives the highest satisfaction is DC, i.e. defect when other cooperates. However, at the system level it is clear that the best combination is CC.

The $\sigma$ profile provides a visualization of information that is already present in payoff matrixes. However,  the outcomes of different actions at different scales are clearer with the $\sigma$ profile, complementing the analysis traditionally carried out in game theory. Moreover, it is easy to include different payoff matrixes at different scales, i.e. when the relationship between satisfaction between scales is non-trivial. Furthermore, the $\sigma$ profile can be used to study not only different actions or strategies, but how changes in the payoff matrixes affect the satisfaction at different scales. This is relevant because e.g. in the complex dynamics of an ecosystem the behavior of some animals or species can change the payoff of other animals or species. These changes are difficult to follow if only matrixes are used.
	
	Note that the scales mentioned so far are spatial, but these can also be temporal, e.g. short term payoffs can be different from long term ones. An example can be given with iterated games: single games are a faster temporal scale, while iterations between the same players can constitute slower temporal scales.
	
\begin{figure}[htbp]
\begin{center}
\includegraphics[width=0.6\textwidth]{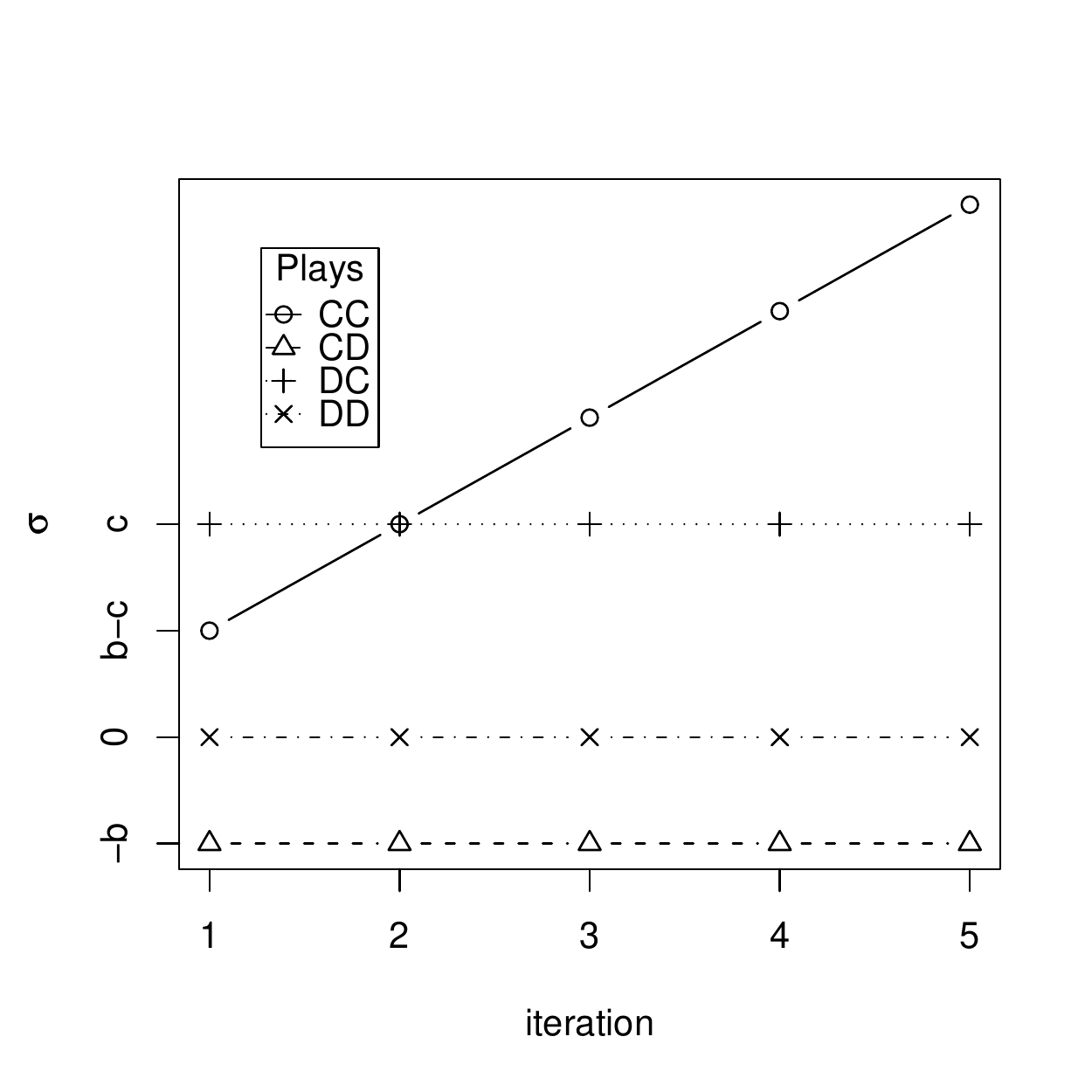}
\caption{$\sigma$ profile for an iterated prisoner's dilemma. At the single game scale, the best play is DC, while after more than two games it is CC. $b=2c$.}
\label{fig:sigmaIPD}
\end{center}
\end{figure}

	Figure \ref{fig:sigmaIPD} shows the temporal $\sigma$ profile for an iterated prisoner's dilemma where players choose between always defect or always cooperate. The following assumption is made: players are able to give a benefit $b$ at a cost $c$ only if their satisfaction is not negative, i.e. if they have enough resources. Thus, the combination DC cannot achieve more than $b$ for agent $A$, since $B$ is left with nothing to give after one game, i.e. $-c$. Like this, with time CC is clearly the best combination \emph{at the individual scale and slow time scale}, since the benefit of defection applies only at the fast time scale.

\section{Nature}

\begin{figure}[htbp]
\begin{center}
\includegraphics[width=0.6\textwidth]{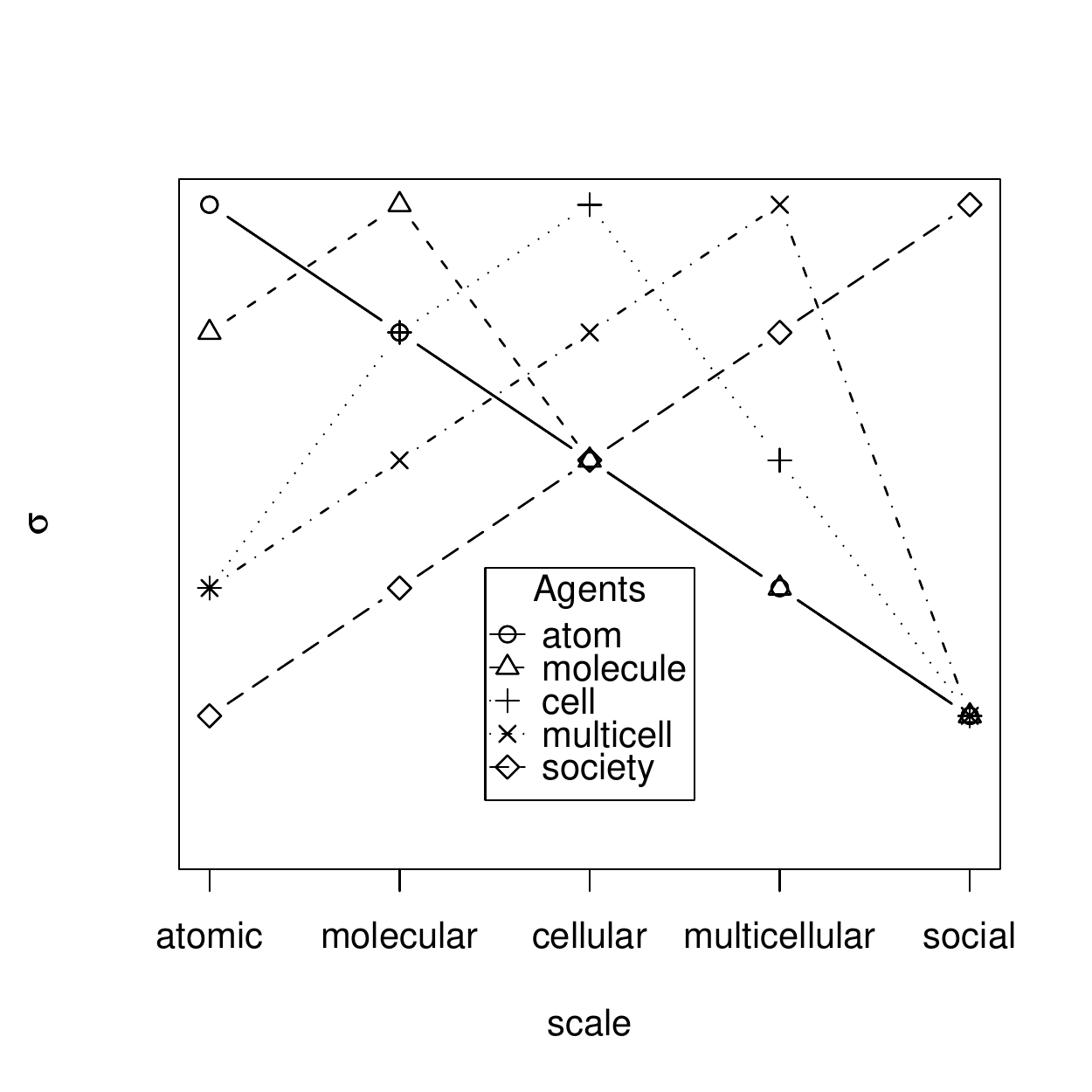}
\caption{$\sigma$ profile across five scales for agents characteristic of each scale.}
\label{fig:sigma}
\end{center}
\end{figure}

Remembering that satisfactions ascribed to agents partially depend on the observer, the $\sigma$ profile can be used to compare satisfactions across scales in nature. Figure \ref{fig:sigma} shows the $\sigma$ profile at five scales, from atomic to social. At the lowest scale, isolated atoms have the highest satisfaction, since they are able to fulfill their goal of reaching a state of minimum energy and highest entropy. When molecules organize atoms, these cannot reach such a satisfactory state as when they are free, so their satisfaction is reduced at the molecular scale. Naturally, isolated molecules have the highest satisfaction at this scale, since they can ``enslave" atoms to reach their goal (minimize chemical potential) and are free from agents of higher scales. Molecules in turn are enslaved by cells, since the latter organize the former to maximize their own satisfaction. The main goals of cells are to survive reproduce. In multicellular organisms, cells are constrained in their reproduction and survival (via apoptosis) to the benefit of the organism. Cancer cells can be seen as ``rebels" to the goals and satisfaction of the organism. Groups and societies also constrain and organize individuals to reach their own goals and increase their satisfaction. The prisoner's dilemma is an example of this last case.

In order for a higher scale structure to maintain itself, its satisfaction has to be greater or equal than that of its components. This leads to an ``enslaving" of the lower scale agents \cite{Haken1988} by the higher scale system, as their satisfaction will be in some cases decreased. However, their survivability will be enhanced, as the system will \emph{mediate} \cite{Michod1997,Michod2000,Michod2003,Heylighen:2006} conflicts between agents to reduce their friction for its own ``selfish" satisfaction. Like this, even when the satisfaction of animals is lower when they are social, they have better chances of survival, since they benefit from the social organization and are able to cope with more environmental complexity as a group. The same applies to cells: they are less ``free" in a multicelular organism, as they obey intracellular signals that can imply their own destruction. Nevertheless, cells within a multicellular organism have more chances to survive than similar ones competing against each other for resources. In a similar way, many molecules would not be able to maintain themselves if it weren't for the organization provided by a cell \cite{Kauffman2008}. Finally, free atoms are more prone to change their minimal energy state by interacting with other atoms than those that are already binded in molecules.

Agents at each scale will try to maximize their own satisfaction, so the only way for higher scale agents to emerge is to mediate or enslave the lower scale agents. The scale dominating the $\sigma$ profile, i.e. with a higher satisfaction, reflects the degree of organization and complexity of the system. In Figure \ref{fig:sigma}, agents characteristic of a scale have the highest satisfaction at that scale. 

\section{Evolution}

There has been much work discussing the evolution of complexity \cite{Bonner1988,BedauEtAl2000,GershensonLenaerts2008}. Our universe has seen an increase of complexity throughout its history, from the Big Bang to the information age. People have described this as an ``arrow" in evolution \cite{Bedau1998,Stewart:2000}. However, some others have seen this increase as a natural drift \cite{McShea:1994,Miconi:2008}. This means that starting with only simple elements, with random variations you can only get more complex. However, this explanation does not account for the increasing speed at which complexity has evolved.

The $\sigma$ profile can be used to gradually measure metasystem transitions \cite{Turchin1977,szathmary97}\footnote{A metasystem transition occurs when the $\sigma$ at a higher scale dominates the profile.}, which clearly indicate an increase of complexity. 
Thus, the $\sigma$ profile can be used to understand better the evolution of complexity. Each agent at its own scale tries to maximize its satisfaction. But a high satisfaction does not always imply a higher evolutionary fitness. Some systems will have high $\sigma$ values at higher or lower scales. But those with high values at high scales will have a higher fitness in comparison, since an agent at a high scale needs to ensure the sustainability and cooperation of all agents at its lower scales to maintain itself. On the other hand, independent agents at lower scales will not be able to do much beyond their own scale to ensure their survivability.

The above scenario does not imply that complexity will always increase. Like with any evolutionary process, a source of variation is needed. Once there is a competition between two different systems, one with a higher organizational scale will tend to win the evolutionary race. Since it is beneficial to have high $\sigma$ at high scales, systems which can evolve higher scales of organization will tend to evolve. And those who can evolve faster will prevail. There are many ways in which cooperation can evolve \cite{Nowak2006}, but this seems to be a general evolutionary tendency, not only in biological systems, but also in economical, technological, and informational systems, where an increasing increase of complexity is also observed  \cite{Kauffman2008}. Whether there is an upper bound for complexity increase is still an open question.

\section{Simulation}

A simple multi-agent simulation was developed in NetLogo \cite{Wilensky1999}.  Agents move randomly in a toroidal virtual world with a certain $step\,size$ at a certain $energy\,cost$. Resources grow randomly at a certain $growth\,rate$. Agents feed on resources, increasing their energy by a certain $resource\,energy$. If the agent's internal $energy \in [0,100]$ reaches zero, it dies. If the $energy$ reaches one hundred, it reproduces by splitting. This reduces the $energy$ of the parent to half, and creates an offspring with the other half of the $energy$.

In this simple simulated ecological system, typical behaviors can be observed depending on the parameters. If the resource $growth\,rate$ or  $resource\,energy$ is low, or the agent's $energy\,cost$ is high or $step\,size$ is small, the agents will become extinct. As these variables change, larger agent populations can be maintained. Depending on the precise variable values, the population sizes of agents and resources can be roughly constant or oscillate around a mean.

To study the benefit of aggregating, a variable $group\,advantage$ is introduced. If an agent is not in an aggregate, every time step its energy is reduced by a certain  $energy\,cost$, as mentioned above. However, if the agent is part of a group, its energy will be reduced by  $energy\,cost/group\,advantage$. Thus, the larger the value of $group\,advantage$, the less energy that an agent in an aggregate will lose. It should be noted, however, that if many agents are aggregated, there will be less resources left for them.

When an agent is born, it is decided with a probability $p_{join}$ whether it will join other agents when it is next to them. Likewise, it is decided with a probability $p_{split}$ whether it will cut links made with or by another agent. These probabilities vary $\pm0.05$ from the parent's probabilities. Depending on the parameter values, agents with different  $p_{join}$ and $p_{split}$ probabilities will have greater advantages, and these will be selected after several generations.

We can measure satisfaction at three levels: the resource level, the agent level, and the system level. The resource $\sigma$ is defined as the percentage of resources available in the environment. It is one if the environment is covered by resources, and zero if there are no resources at all. The agent $\sigma$ is measured with their $energy/100$. It is one if the agents are about to reproduce, and zero if they are dead. The system $\sigma$ is defined as the proportion of agents that are joined in the largest group. It is one if all the agents are joined in one group, and zero if no agent is joined.

\begin{figure}[htbp]
\begin{center}
\includegraphics[width=.8\textwidth]{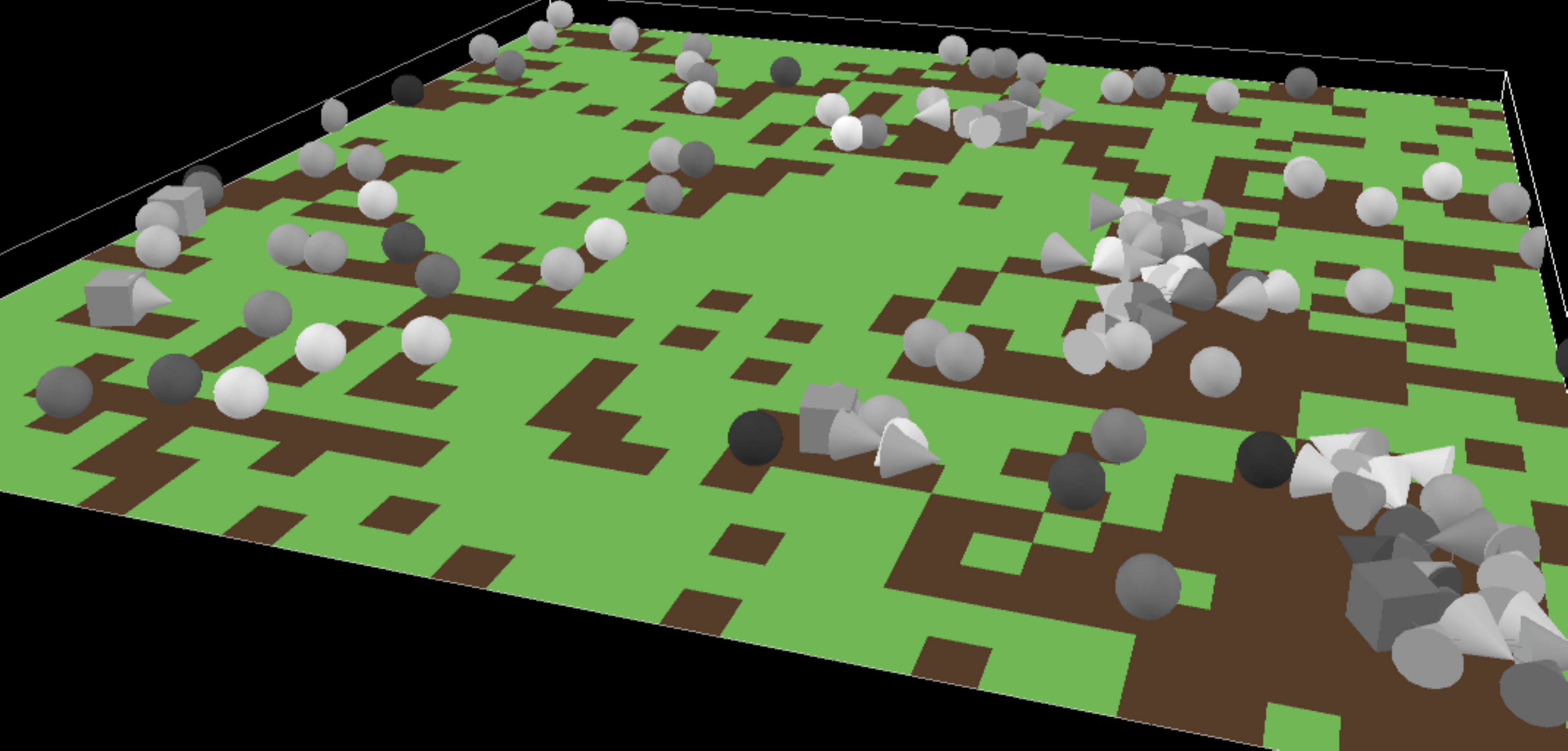}
\caption{Screenshot of simulation. Green patches contain resource, dark brown ones are empty. Lighter agents have more energy than darker ones. Independent agents are represented by spheres. In groups, the movements of a cube agent are followed by cone agents. Areas nearby agent groups are scarcer in resources.}
\label{fig:screenshot}
\end{center}
\end{figure}

A screenshot of the simulation can be seen in Fig. \ref{fig:screenshot}. The reader is invited to try the simulation at the URL \url{http://homepages.vub.ac.be/~cgershen/MO/MO.html}. 

\section{Experiments}

\begin{figure}%[htp]
     \centering
     \subfigure{
          \label{fig:resourceS}
          \includegraphics[width=.4\textwidth]{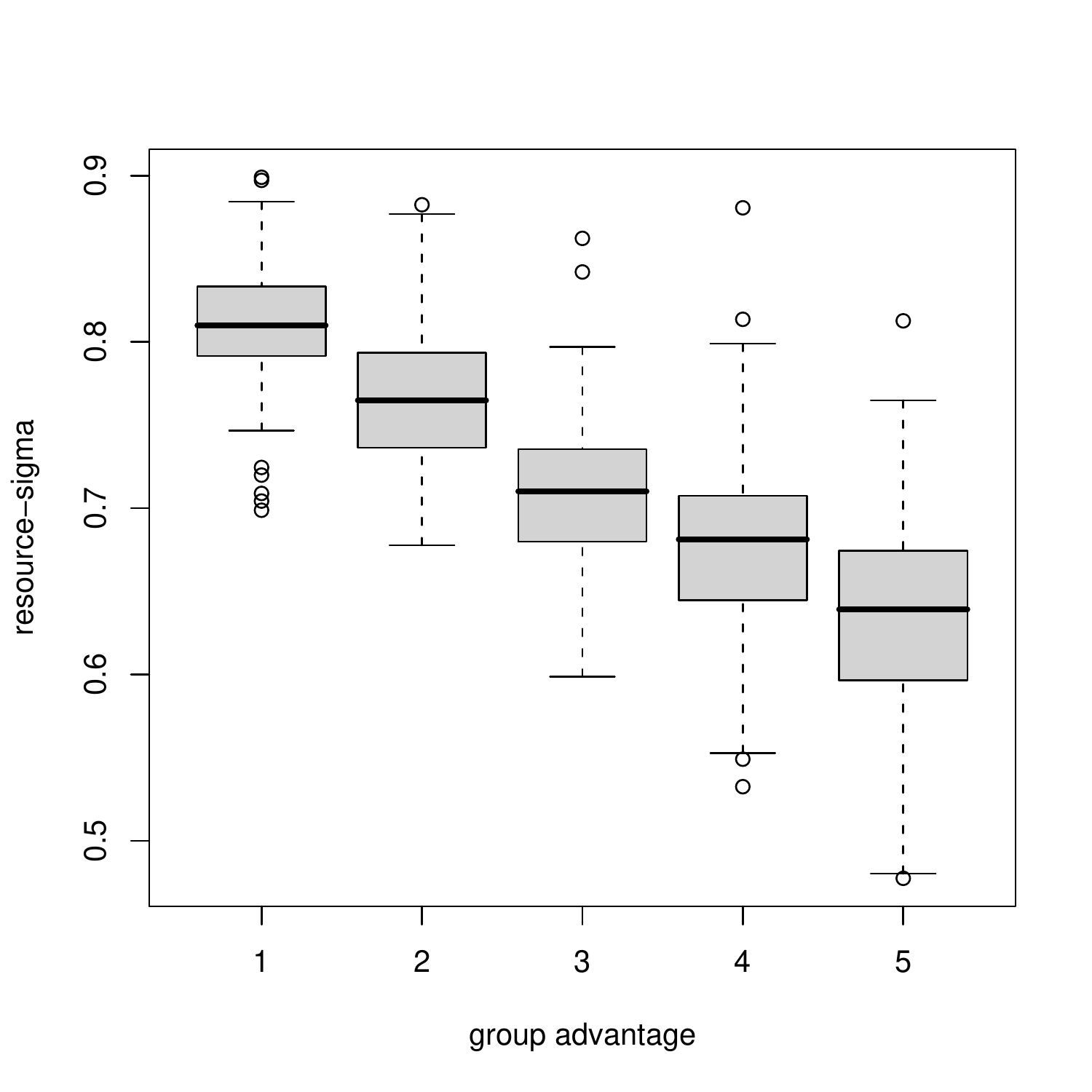}}
     %\hspace{.3in}
     \subfigure{
          \label{fig:agentS}
          \includegraphics[width=.4\textwidth]{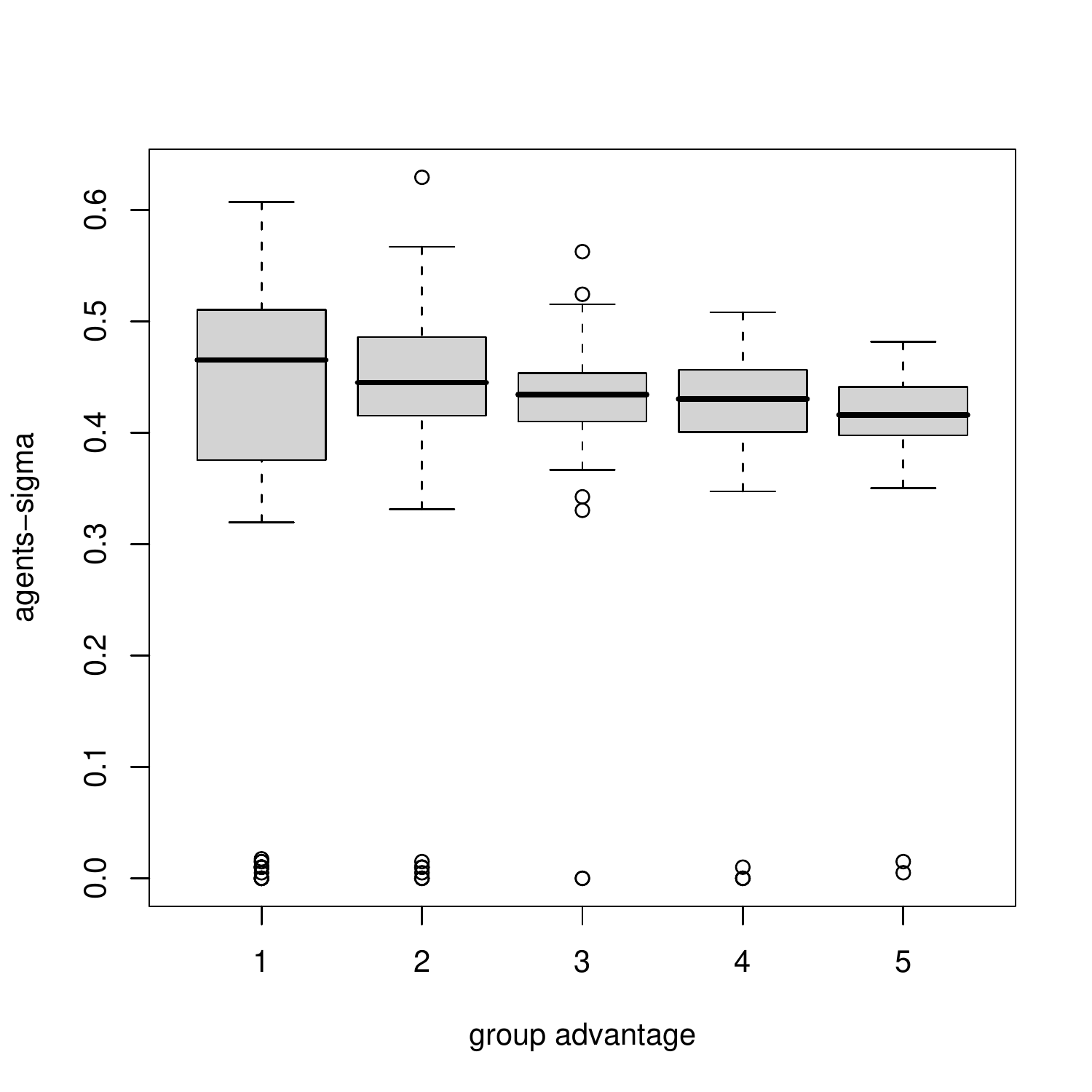}}\\
     %\vspace{.1in}
     \subfigure{
           \label{fig:systemS}
           \includegraphics[width=.4\textwidth]{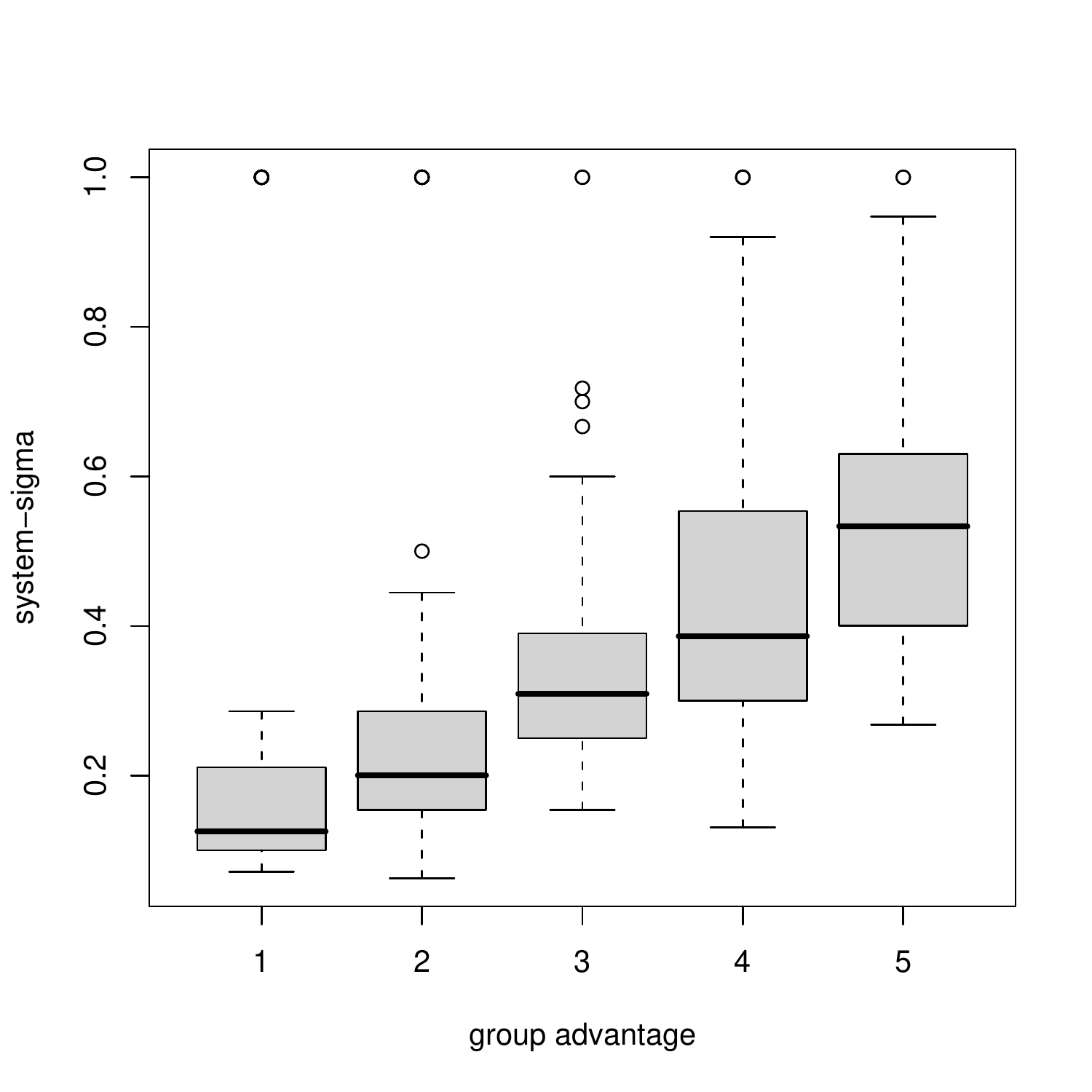}}
     %\hspace{.3in}
     \subfigure{
           \label{fig:agentP}
          \includegraphics[width=.4\textwidth]{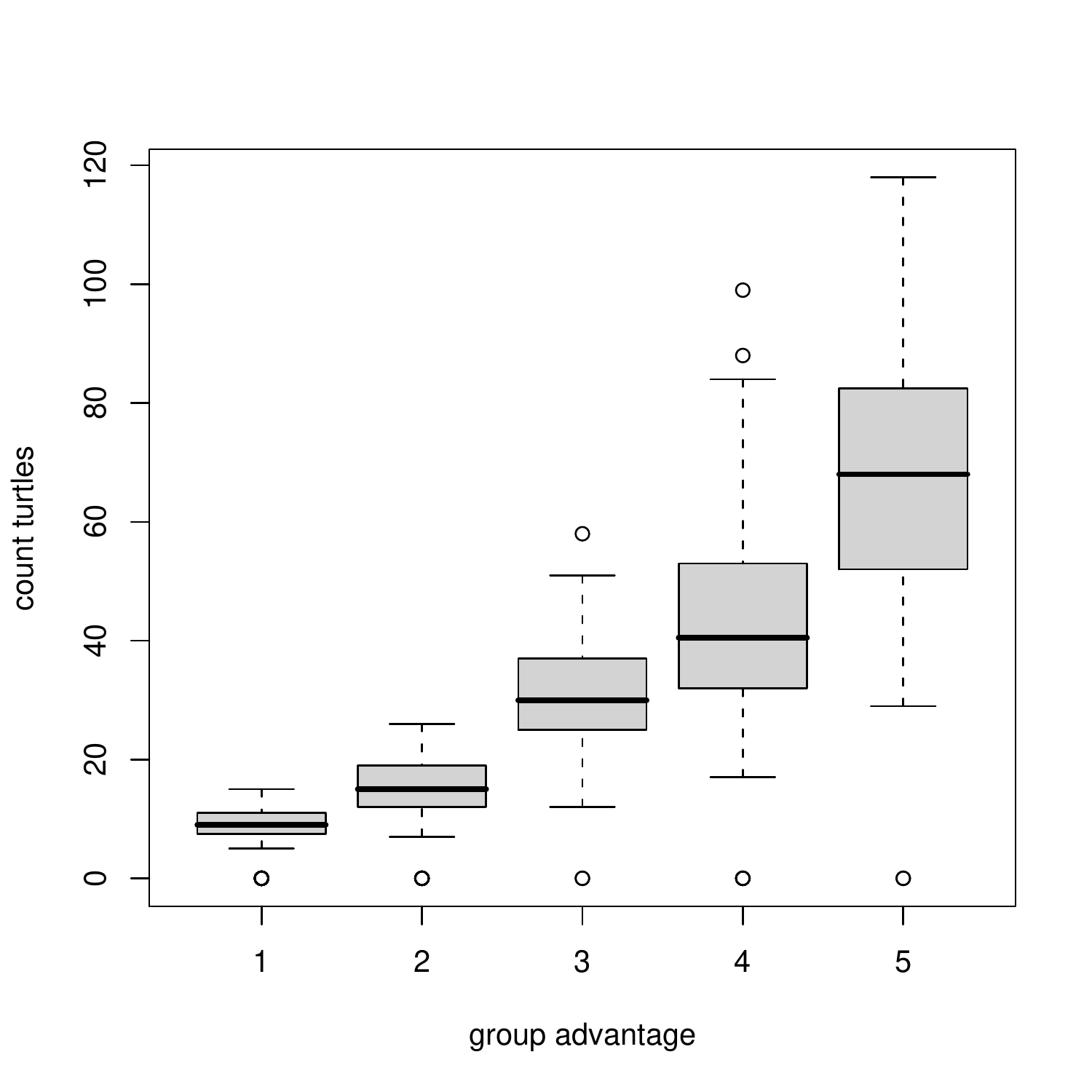}}\\
     %\vspace{.1in}
     \subfigure{
           \label{fig:pj}
           \includegraphics[width=.4\textwidth]{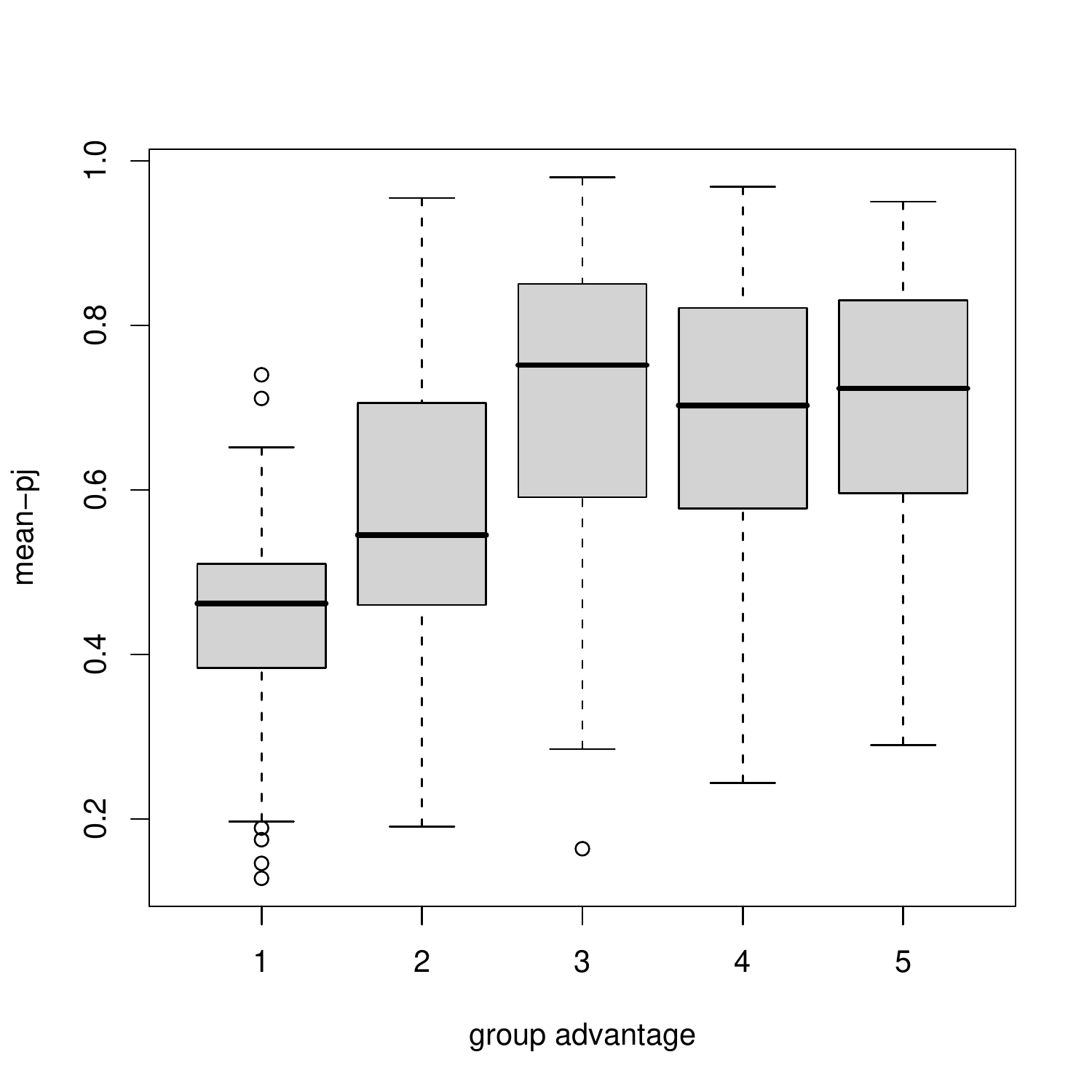}}
     %\hspace{.3in}
     \subfigure{
           \label{fig:ps}
          \includegraphics[width=.4\textwidth]{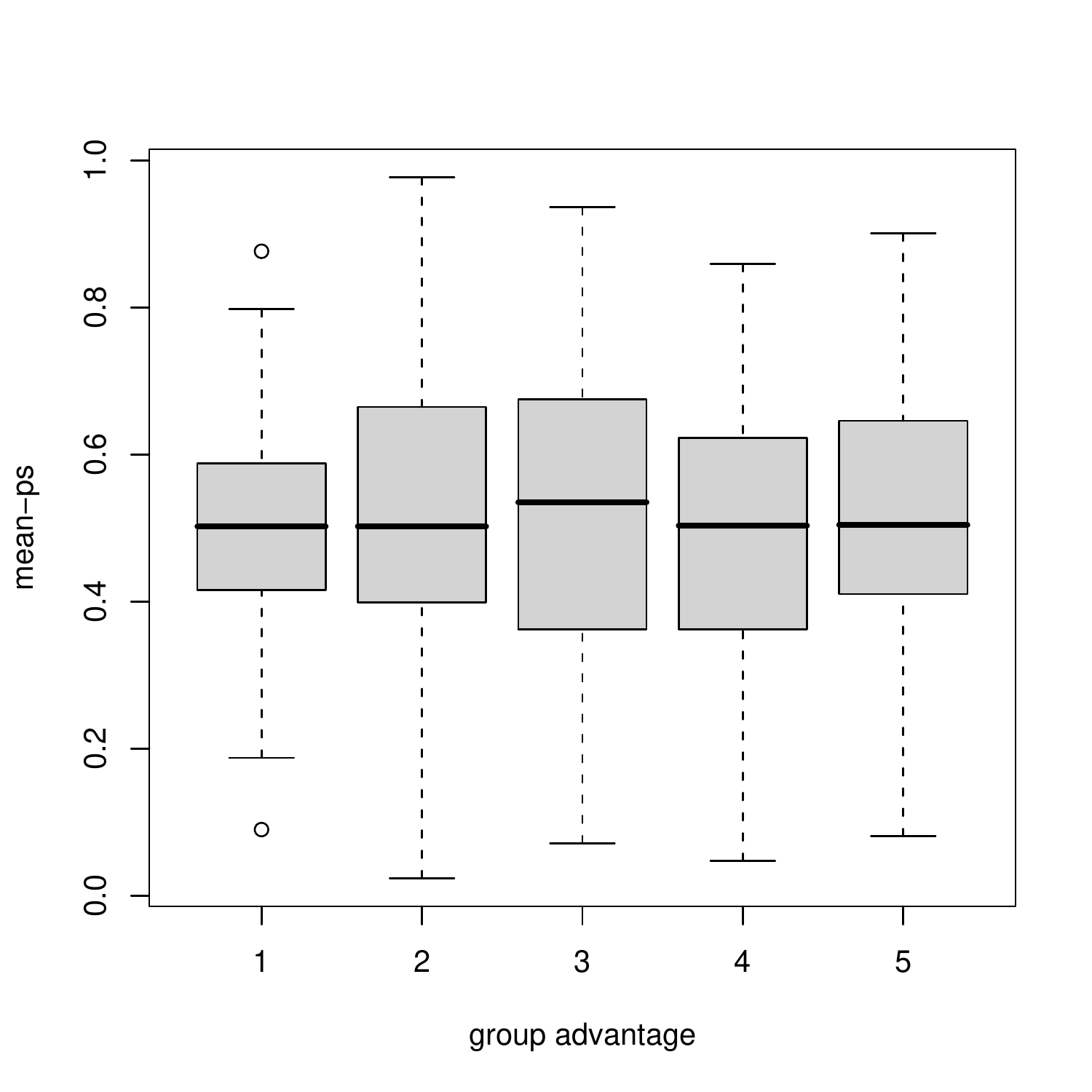}}
     \caption{Simulation results as the $group advantage$ is increased: (a) Resource $\sigma$, (b) Agent $\sigma$, (c) System $\sigma$, (d) Agent population, (e) mean $p_{join}$, and (e) mean $p_{split}$. } 
     \label{fig:results}
\end{figure}

Figure \ref{fig:results} shows results of one hundred simulation runs of ten thousand time steps for different values of $group\,advantage$. All simulations start with an initial population of one hundred agents with $p_{join}=p_{split}=0.5$  and a random $energy$. Table \ref{table:parameters} lists the values of parameters  used.

\begin{table}[htdp]

\begin{center}
\begin{tabular}{ll}
\hline
Variable & Value\\
\hline
$resource\,energy$ & 10\\
$resource\,growth\,rate$ & 0.1\\
$energy\,cost$ & 2 \\
$step\,size$ & 0.2\\
\hline
\end{tabular}
\end{center}
\label{table:parameters}
\caption{Parameter values used in simulation experiments.}
\end{table}%

We can see that as $group\,advantage$ is increased, the system $\sigma$ also increases (Fig. \ref{fig:systemS}). On the other hand, the resource $\sigma$ and agent $\sigma$ are reduced (Figs. \ref{fig:resourceS} and \ref{fig:agentS} ). However, the survivability of the individual agents is increased, as indicated by the population size (Fig. \ref{fig:agentP}).

For higher values of $group\,advantage$, there is a selective pressure towards joining in groups, so the mean $p_{join}$ is increased (Fig. \ref{fig:pj}). This is not favored for low values of $group\,advantage$, since close agents need to share local resources, leading to friction between neighbors. For this case, it is more advantageous for the agents to spread as much as possible in their environment, to avoid the friction. On the other hand, high values of $group\,advantage$ reduce this friction and promote the agent aggregation.
Since a high $p_{join}$ will make agents to join groups even if they split constantly from them, there is no pressure on the value of $p_{split}$ (Fig. \ref{fig:pj}). Note that the reproduction and mutation takes place at the agent level, so there is no direct selection of systems. However, the properties of a high system $\sigma$ give better chances of survival to agents, even if their $\sigma$ is lower compared to the case when the system $\sigma$ is low.

These experiments are intended to illustrate the concepts presented in the paper. They are not attempted as a proof. Concepts can only prove their usefulness and suitability with time.

\section{Conclusions}

The $\sigma$ profile has several potential uses to describe and compare systems at multiple scales. One explored here is the difference between satisfaction (payoff) and survivability. A high satisfaction does not imply survivability. This can seem as a problem for some game theoretical formalizations. However, as we observe the satisfactions of agents at different scales (spatial and temporal), it is clear that the survivability of a system is related to the satisfaction of the highest scale.

Systems can achieve high satisfaction with \emph{mediators} \cite{Michod1997,Michod2000,Michod2003,Heylighen:2006} to reduce friction between agents. Friction reduction can be seen as a generalization of cooperation, which is essential in the emergence of new levels of organization \cite[p. 1563]{Michod2003,Nowak2006}.

%\section{Acknowledgements}

%I should like to thank ***. 
%This work was supported by the FWO, Belgium.

%\bibliographystyle{cgg}
\bibliographystyle{unsrt}

\bibliography{carlos,sos,complex,evolution,traffic}

\begin{thebibliography}{10}

\bibitem{GershensonDCSOS}
Carlos Gershenson.
\newblock {\em Design and Control of Self-organizing Systems}.
\newblock CopIt Arxives, Mexico, 2007.
\newblock http://copit-arxives.org/TS0002EN/TS0002EN.html.

\bibitem{Maes1994}
Pattie Maes.
\newblock Modeling adaptive autonomous agents.
\newblock {\em Artificial Life}, 1(1\&2):135--162, 1994.

\bibitem{WooldridgeJennings1995}
M.~Wooldridge and N.~R~. Jennings.
\newblock Intelligent agents: Theory and practice.
\newblock {\em The Knowledge Engineering Review}, 10(2):115--152, 1995.

\bibitem{Wooldridge2002}
Michael Wooldridge.
\newblock {\em An Introduction to {MultiAgent} Systems}.
\newblock John Wiley and Sons, Chichester, England, 2002.

\bibitem{Schweitzer2003}
Frank Schweitzer.
\newblock {\em Brownian Agents and Active Particles. Collective Dynamics in the
  Natural and Social Sciences}.
\newblock Springer Series in Synergetics. Springer, Berlin, 2003.

\bibitem{Ashby1962}
W.~Ross Ashby.
\newblock Principles of the self-organizing system.
\newblock In H.~Von Foerster and G.~W. {Zopf, Jr.}, editors, {\em Principles of
  Self-Organization}, pages 255--278, Oxford, 1962. Pergamon.

\bibitem{GershensonHeylighen2003a}
Carlos Gershenson and Francis Heylighen.
\newblock When can we call a system self-organizing?
\newblock In W~Banzhaf, T.~Christaller, P.~Dittrich, J.~T. Kim, and J.~Ziegler,
  editors, {\em Advances in Artificial Life, 7th European Conference, {ECAL}
  2003 {LNAI} 2801}, pages 606--614, Berlin, 2003. Springer.

\bibitem{vonNeumannMorgenstern1944}
John von Neumann and Oskar Morgenstern.
\newblock {\em Theory of Games and Economic Behavior}.
\newblock Princeton University Press, 1944.

\bibitem{JMS1982}
John Maynard~Smith.
\newblock {\em Evolution and the Theory of Games}.
\newblock Cambridge University Press, 1982.

\bibitem{Tucker:1950}
Albert Tucker.
\newblock A two-person dilemma.
\newblock Reprinted in UMAP Journal 1 (1980), 101., 1950.

\bibitem{Poundstone:1992}
W~Poundstone.
\newblock {\em Prisoner's Dilemma: John Von Neumann, Game Theory and the Puzzle
  of the Bomb}.
\newblock Doubleday, New York, NY, USA, 1992.

\bibitem{Axelrod1984}
R.~M. Axelrod.
\newblock {\em The Evolution of Cooperation}.
\newblock Basic Books, New York, 1984.

\bibitem{Nowak2006}
Martin~A. Nowak.
\newblock Five rules for the evolution of cooperation.
\newblock {\em Science}, 314:1560--1563, December 2006.

\bibitem{BarYam2004}
Y.~Bar-Yam.
\newblock Multiscale variety in complex systems.
\newblock {\em Complexity}, 9(4):37--45, 2004.

\bibitem{Prokopenko:2008}
Mikhail Prokopenko, Fabo Boschetti, and Alex Ryan.
\newblock An information-theoretic primer on complexity, self-organisation and
  emergence.
\newblock {\em Complexity}, In Press.

\bibitem{Ashby1956}
W.~Ross Ashby.
\newblock {\em An Introduction to Cybernetics}.
\newblock Chapman \& Hall, London, 1956.

\bibitem{Haken1988}
Hermann Haken.
\newblock {\em Information and Self-organization: A Macroscopic Approach to
  Complex Systems}.
\newblock Springer-Verlag, Berlin, 1988.

\bibitem{Michod1997}
Richard~E. Michod.
\newblock Cooperation and conflict in the evolution of individuality. i.
  multi-level selection of the organism.
\newblock {\em American Naturalist}, 149:607--645, 1997.

\bibitem{Michod2000}
Richard~E. Michod.
\newblock {\em Darwinian Dynamics: Evolutionary Transitions in Fitness and
  Individuality}.
\newblock Princeton University Press, Princeton, NJ, 2000.

\bibitem{Michod2003}
Richard~E. Michod.
\newblock Cooperation and conflict mediation during the origin of
  multicellularity.
\newblock In P.~Hammerstein, editor, {\em Genetic and Cultural Evolution of
  Cooperation}, chapter~16, pages 261--307. MIT Press, Cambridge, MA, 2003.

\bibitem{Heylighen:2006}
Francis Heylighen.
\newblock Mediator evolution: a general scenario for the origin of dynamical
  hierarchies.
\newblock In Diederik Aerts, Bart {D'Hooghe}, and Nicole Note, editors, {\em
  Worldviews, Science and Us}. World Scientific, 2006.

\bibitem{Kauffman2008}
Stuart~A. Kauffman.
\newblock {\em Reinventing the Sacred: A New View of Science, Reason, and
  Religion}.
\newblock Basic Books, 2008.

\bibitem{Bonner1988}
John~Tyler Bonner.
\newblock {\em The Evolution of Complexity, by Means of Natural Selection}.
\newblock Princeton University Press, 1988.

\bibitem{BedauEtAl2000}
M.~Bedau, J.~McCaskill, P.~Packard, S.~Rasmussen, D.~Green, T.~Ikegami,
  K.~Kaneko, and T.~Ray.
\newblock {Open Problems in Artificial Life}.
\newblock {\em Artificial Life}, 6(4):363--376, 2000.

\bibitem{GershensonLenaerts2008}
Carlos Gershenson and Tom Lenaerts.
\newblock Evolution of complexity.
\newblock {\em Artificial Life}, 14(3):1--3, Summer 2008.
\newblock Special Issue on the Evolution of Complexity.

\bibitem{Bedau1998}
Mark~A. Bedau.
\newblock Four puzzles about life.
\newblock {\em Artificial Life}, 4:125--140, 1998.

\bibitem{Stewart:2000}
John Stewart.
\newblock {\em Evolution's Arrow: The Direction of Evolution and the Future of
  Humanity}.
\newblock Chapman Press, 2000.

\bibitem{McShea:1994}
D.~McShea.
\newblock Mechanisms of large-scale evolutionary trends.
\newblock {\em Evolution}, 48:1747---1763, 1994.

\bibitem{Miconi:2008}
T~Miconi.
\newblock Evolution and complexity: the double-edged sword.
\newblock {\em Artificial Life}, 14(3), Summer 2008.
\newblock Special Issue on the Evolution of Complexity.

\bibitem{Turchin1977}
Valentin Turchin.
\newblock {\em The Phenomenon of Science. A Cybernetic Approach to Human
  Evolution}.
\newblock Columbia University Press, New York, 1977.

\bibitem{szathmary97}
J.~Maynard Smith and E.~Szathm{\'a}ry.
\newblock {\em The major transitions in evolution}.
\newblock Oxford University Press, 1995.

\bibitem{Wilensky1999}
Uri Wilensky.
\newblock {NetLogo}, 1999.

\end{thebibliography}

\end{document}